\documentclass[twocolumn,showpacs,preprintnumbers,amsmath,amssymb]{revtex4}

\usepackage{graphicx}
\usepackage{dcolumn}
\usepackage{bm}

\begin{document}

\preprint{}

\title{Scale-free topology of the interlanguage links in Wikipedia}

\author{\L{}ukasz Bolikowski}%
 \email{bolo@icm.edu.pl}
\affiliation{%
Interdisciplinary Center for Mathematical and Computational Modeling, \\
University of Warsaw, Poland
}%

\date{\today}

\begin{abstract}
The interlanguage links in Wikipedia
connect pages on the same subject written in different languages.
In theory, each connected component should be a clique and cover one topic.
However, incoherent edits and obvious mistakes result in topic coalescence,
yielding a non-trivial topology that is studied in this paper.
We show that the component size distribution obeys the power law,
and we explain anomalies in the distribution
as results of certain edit conventions.
Next, we propose a method of filtering out the cliques and
study basic properties of the resulting skeleton, which turns out to be scale-free.
\end{abstract}

\maketitle

In the recent years
Wikipedia has been increasingly a subject of scientific study, both qualitative and quantitative
\footnote{Wikipedia maintains a list of conference presentations and peer-reviewed papers
that focus on the project, see: {\small\tt http://en.wikipedia.org/wiki/WP:ACST}}.
Its content serves as an excellent example of a large complex network \cite{PhysRevE:2006:Zlatic},
which exhibits exponential growth in the number of contributors and text content \cite{ICWSM:2007:Almeida}.
The growth has been described in terms of the preferential attachment mechanism \cite{PhysRevE:2006:Capocci},
and the dynamics of user contributions (e.g. conflict patterns) have been thoroughly studied
\cite{VAST:2007:Suh, InfVis:2008:Brandes}.

In this paper we examine yet another, so far undescribed, facet of this network:
the topology of the interlanguage links.
The analysis is based on database dumps retrieved on August 27, 2008.
At that time, the interlanguage links were defined as
``links from any page (most notably articles) in one Wikipedia language
to the same subject in another Wikipedia language''
\cite{Wikipedia:Help:InterlanguageLinks}.
Given this definition, the {\em expected} topology of the network is trivial:
each subject should be represented by a separate, isolated clique consisting of all the pages on the subject,
each clique should contain at most one page from any given language edition,
and there should be no other links in the network.
Mathematically speaking, the sum of all the links should form an equivalence relation $\equiv$,
satisfying an additional condition:
\begin{equation}
a \equiv b \Rightarrow a = b \vee lang(a) \neq lang(b)
\end{equation}

However, the software engine that powers Wikipedia
does not enforce coherence of the network:
each page maintains a list of outgoing interlanguage links.
There are user-controlled programs, so called bots,
which add the {\em missing} links by performing symmetric and transitive closure.
This means that for each link $a \to b$ the bots add $b \gets a$ if missing,
and for each pair of links $a \to b \to c$ add $a \to c$ if missing.
The opposite problem of removing the {\em extra} links is not trivial:
while it is easy to detect a conflict using an automaton, resolving it
requires understanding of the contents of the involved pages.
For example, a simple program traversing the network may discover
the following conflict (a real example, which has already been corrected):
{\em en:Tap (valve)} $\equiv$
{\em it:Rubinetto} $\equiv$
{\em es:Grifo} $\equiv$
{\em en:Griffin}.
Here a program may raise a flag since two pages in the same language are present in one connected component,
but a human will have to read the articles to find the incorrect link(s).

Let us proceed to study the properties of two networks of interlanguage links:
one connecting the articles ($\mathcal{A}$),
and the other connecting the categories ($\mathcal{C}$).
In both cases we will treat the networks as undirected graphs,
assuming a link $a - b$ iff there is an interlanguage link $a \to b$ or $b \gets a$.

Network $\mathcal{A}$ consists of 11\,510\,142 nodes and 89\,339\,694 links.
Approx. 42\% of the nodes are isolated, and the remaining nodes are
grouped into 1\,223\,183 connected components.
Network $\mathcal{C}$ consists of 1\,724\,088 nodes and 13\,902\,852 links,
approx. 51.5\% of the nodes are isolated, and the rest are grouped
into 118\,039 connected components.

We will say that a connected component is {\em coherent} when no two pages
are in the same language 
\footnote{We use the terms ``node'' and ``page'' interchangeably.
Both are equivalent to ``article'' in the context of network $\mathcal{A}$,
and to ``category'' in the context of network $\mathcal{C}$.},
and that it is {\em complete} when it contains all the possible links (i.e., is a clique).
There are 59\,323 incoherent components in $\mathcal{A}$ and 6\,152 incoherent components in $\mathcal{C}$.
In both cases it is approx. 5\% of all the non-singleton connected components.
Completeness is correlated with coherence: for example in the case of $\mathcal{A}$,
63\% of the coherent components are complete,
and 99\% contain at least half of all the possible links.
On the other hand, none of the incoherent components are complete,
and only about 61\% contain at least half of all the possible links.

\begin{figure*}
\begin{center}$
\begin{array}{ccc}
\includegraphics[width=0.29\textwidth]{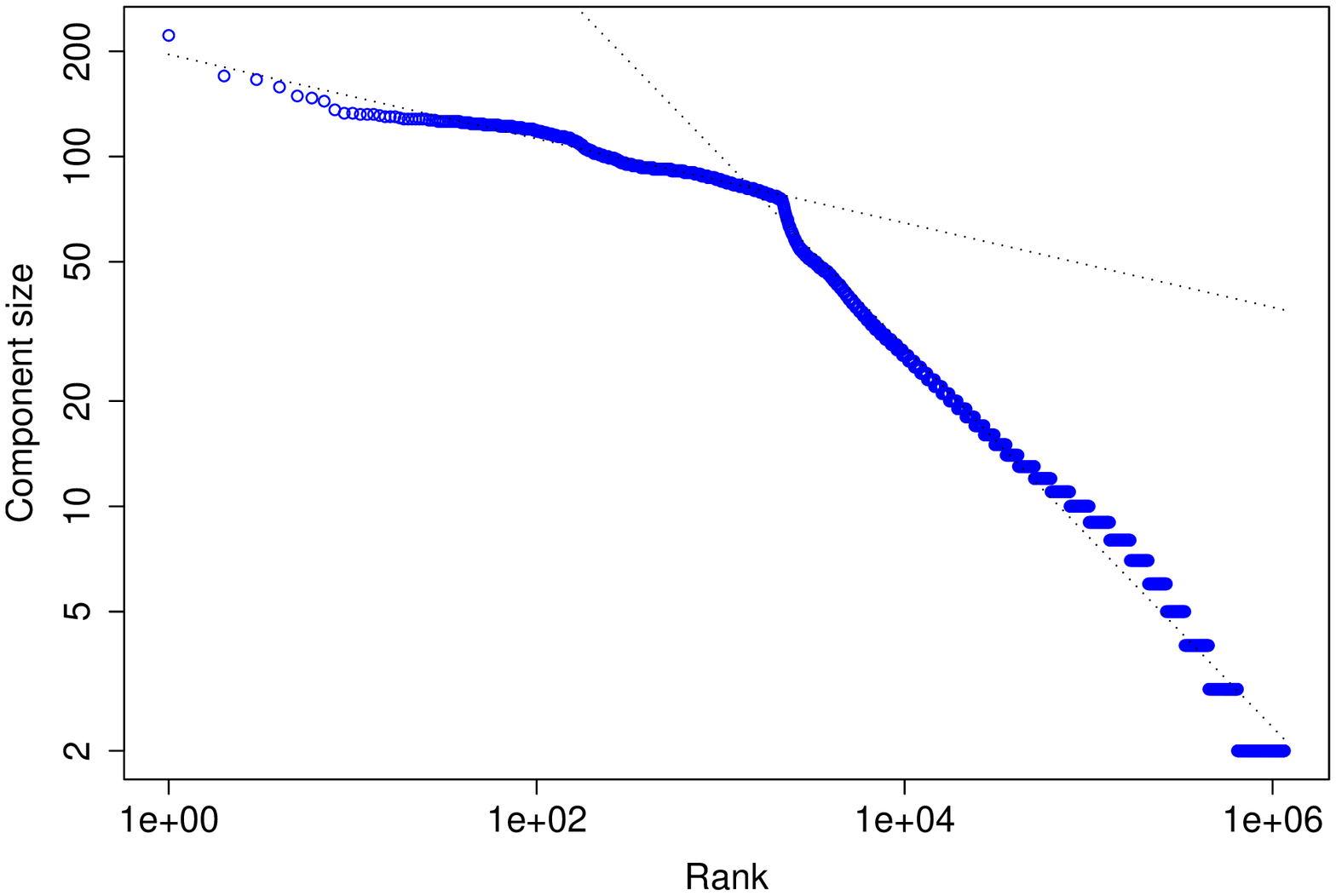} &
\includegraphics[width=0.29\textwidth]{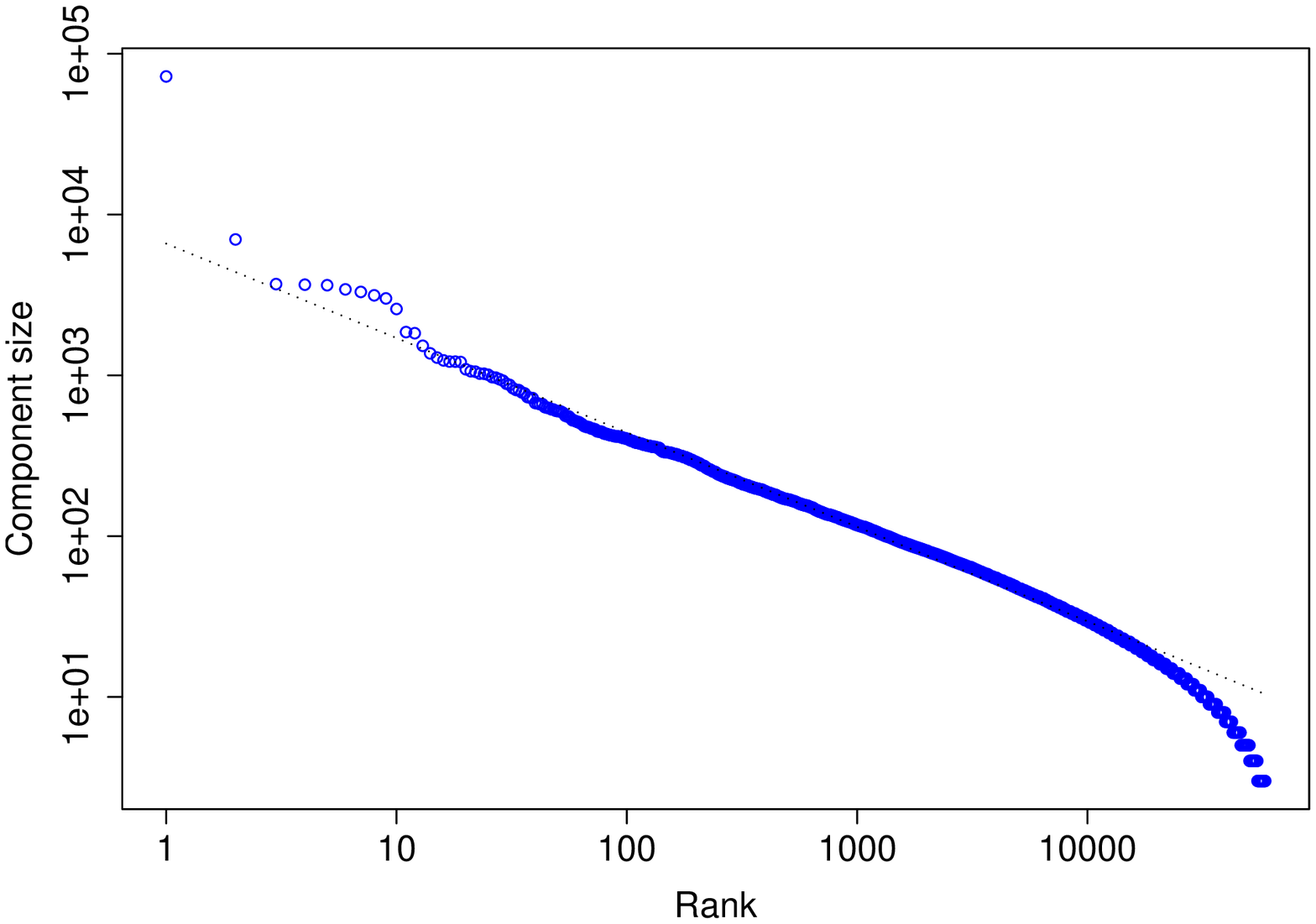} &
\includegraphics[width=0.29\textwidth]{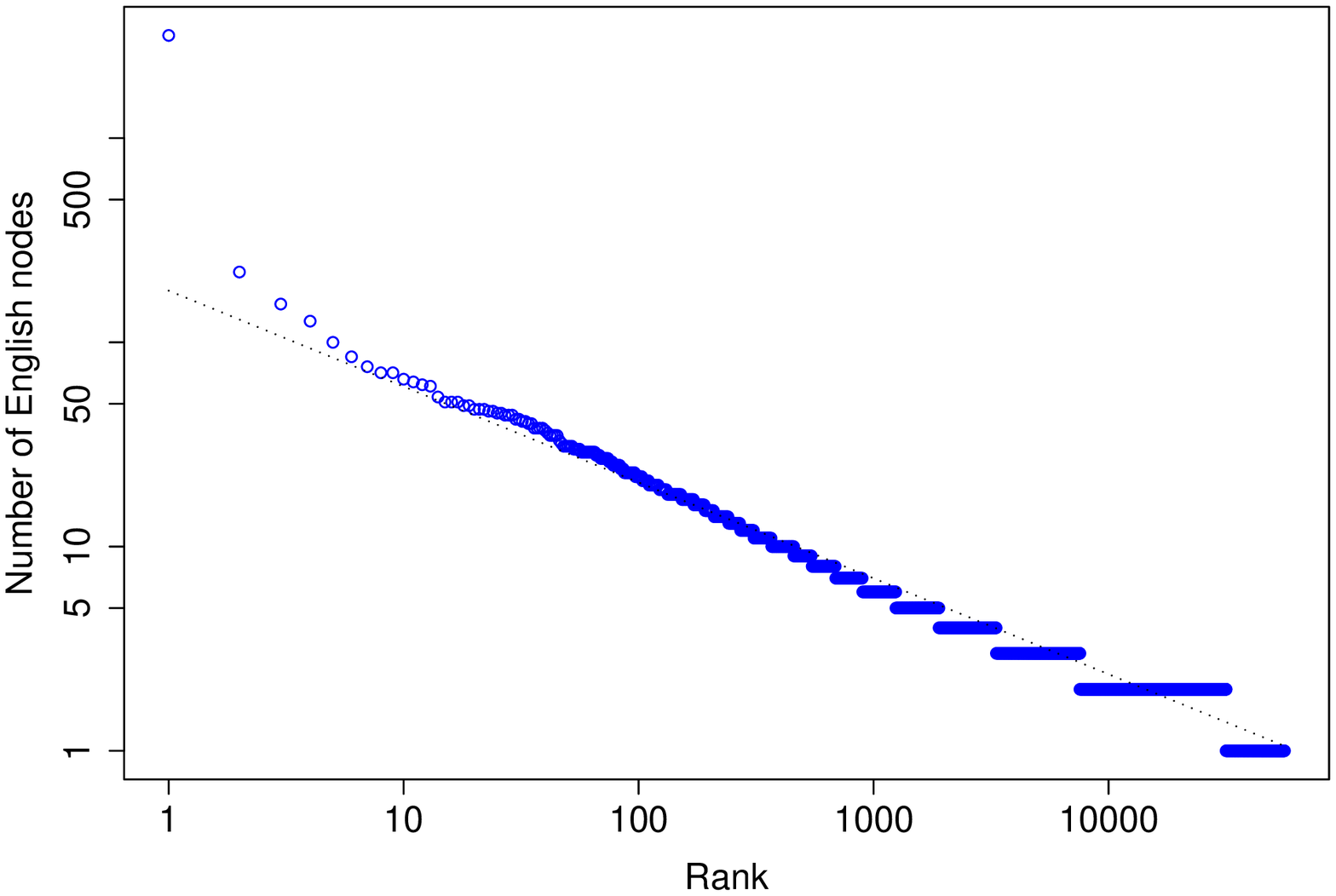} \\
\includegraphics[width=0.29\textwidth]{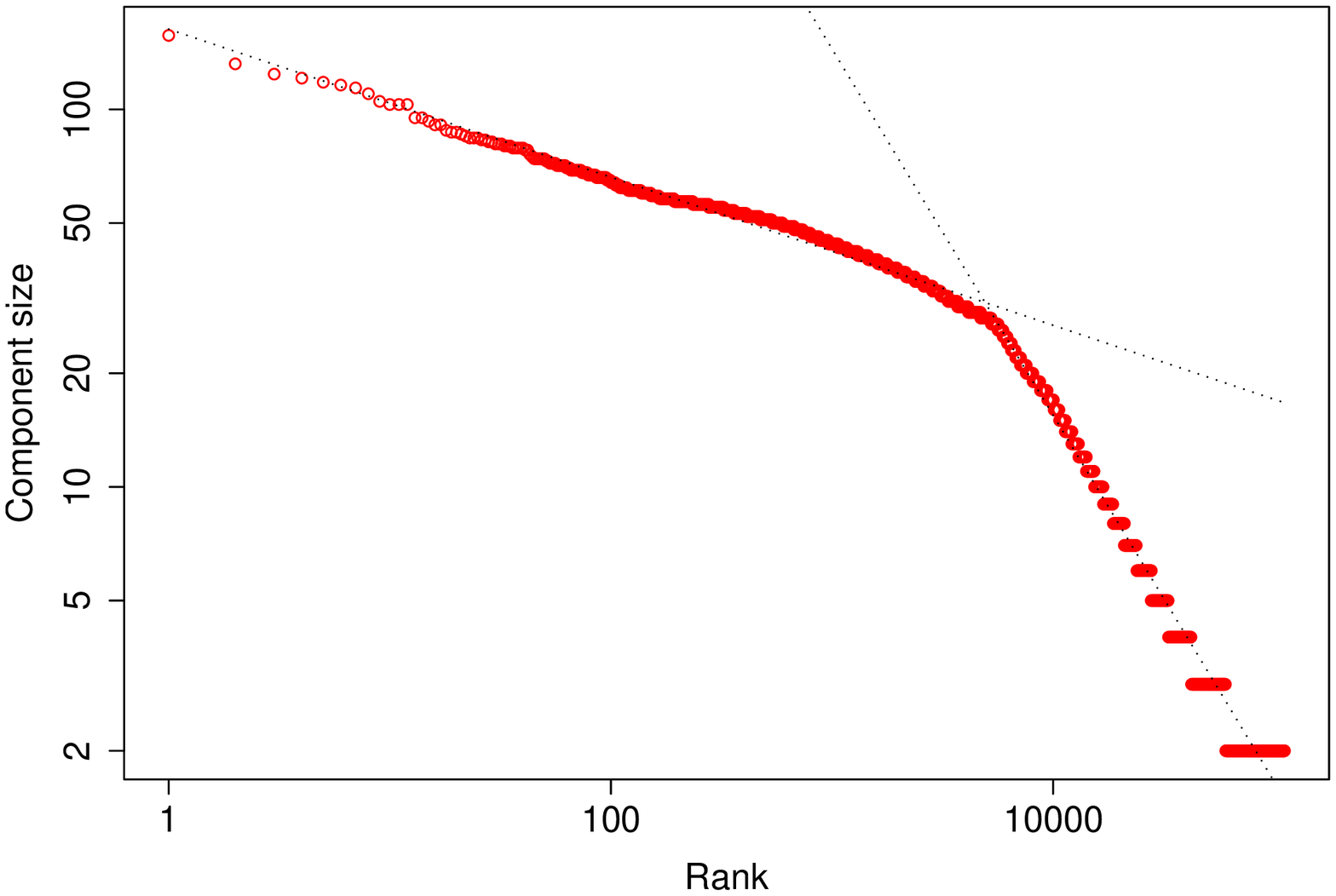} &
\includegraphics[width=0.29\textwidth]{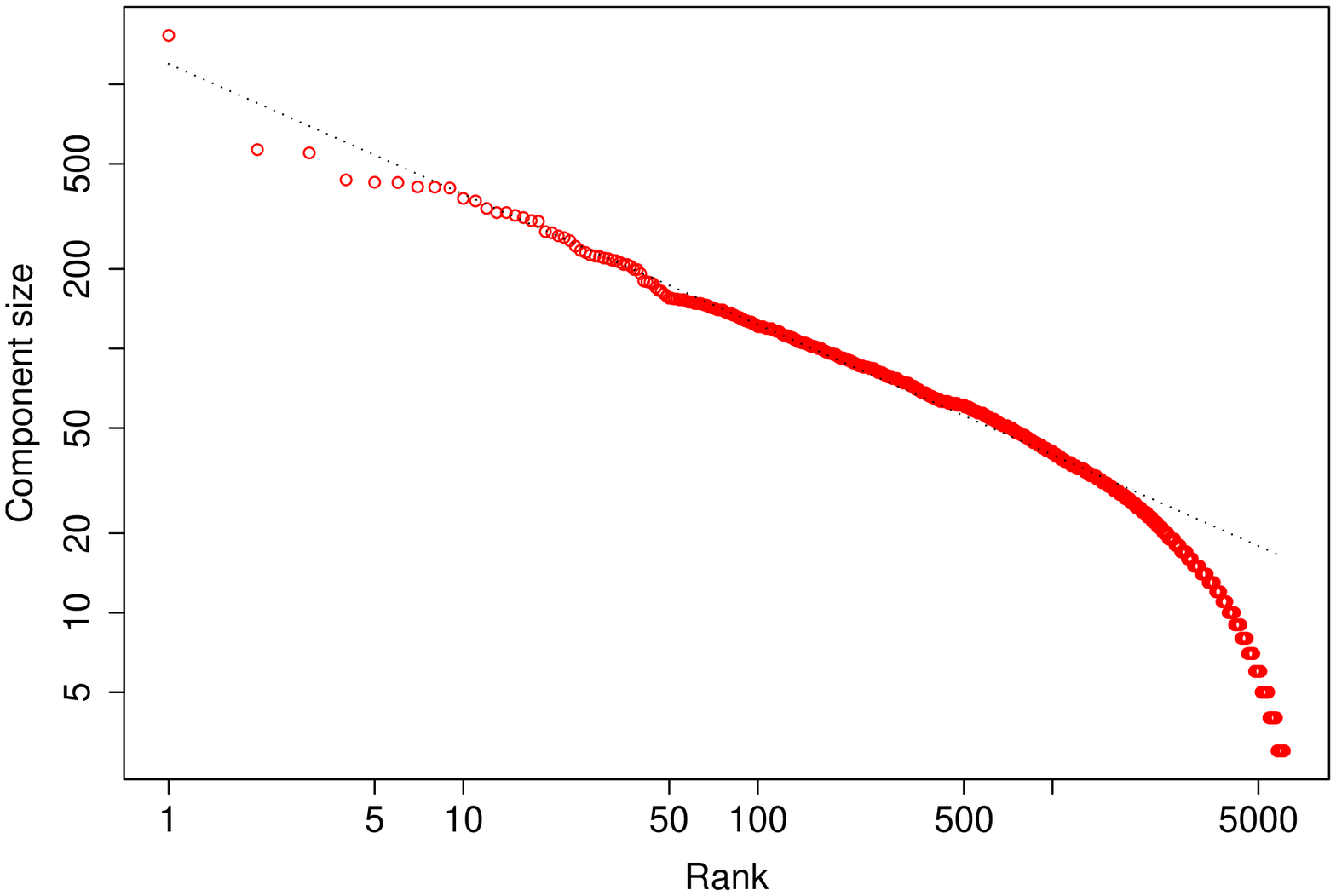} &
\includegraphics[width=0.29\textwidth]{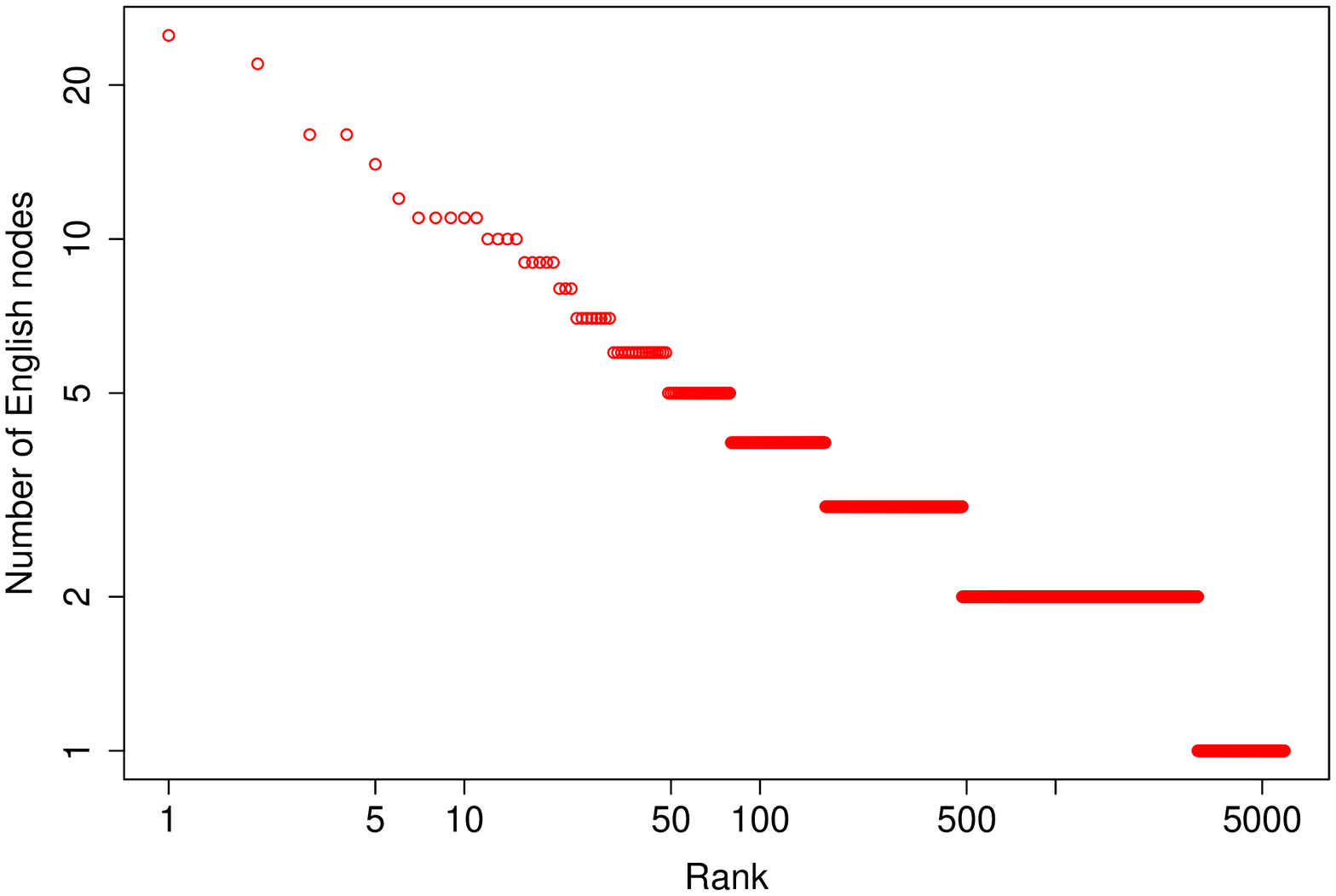}
\end{array}$
\end{center}
\caption{\label{fig:compsizes} Component sizes for the article network (top row)
and the category network (bottom row).
Sizes of the coherent components (left column),
the incoherent components (middle column)
and the number of English pages in the incoherent components (right column)
are all plotted against their ranks.
Each plot has a log-log scale.
The ``king effect'' \cite{ComptesRendus:1996:Laherrere} in the case of incoherent components
and the influence of mass-produced date-related topics on the shapes of the coherent component distributions
are visible (both features are discussed in the text).}
\end{figure*}

Component size vs.\! rank plots for $\mathcal{A}$ and $\mathcal{C}$ are
presented in Figure \ref{fig:compsizes}. 
Values for the coherent and incoherent components are plotted separately.
For the incoherent components of each network,
the number of English pages (a coarse measure of the number of topics) vs. rank is also shown.
All the plots have logarithmic scales.

The plotted points are (piecewise) well approximated by straight lines,
which indicates that a component's size $s$ is a power-law function
of its rank $r$, namely: $s \sim r^{-\gamma}$.
Power-law distributions are encountered in diverse settings,
for example:
the distribution of city sizes \cite{PhysRevLett:1998:Marsili},
occurrences of DNA base pair sequences \cite{PhysRevLett:1994:Mantegna},
and number of sent e-mails \cite{PhysRevE:2002:Ebel},
all follow the power law.
For an excellent description of the distribution
and numerous examples of its occurrences, see Ref. \cite{ContPhys:2005:Newman}
and \cite{arXiv:2007:Clauset}.

Let us take a closer look at the obtained distributions.
In the case of the coherent components of $\mathcal{A}$
(top left panel of Figure \ref{fig:compsizes}),
there are two clear regimes.
Most of the top 2\,200 or so components (the first regime),
each covered by at least 75 language editions,
contain articles on the years of the current
era and centuries.
Such articles are easy to create in an automated way,
and it is easy to maintain the interlanguage links to corresponding articles
in the other language editions
(easy maintenance explains why the components are coherent).
An informal competition among the language editions
for the largest number of articles might be an additional motivation
for the mass-creation of the date-related pages.

Similarly, two regimes in the component size distribution of $\mathcal{C}$
(bottom left panel of Figure \ref{fig:compsizes}) can also be observed,
although the transition between them is smoother than in the previous case.
Date-related topics account for about 75\% of the top
5\,000-6\,000 components (each containing at least 26-28 nodes).
Among these are categories for years, decades, centuries,
births and deaths in a given year, and (for the recent times)
films and video games in a given year.
Other prominent categories are: countries (including ``History of \ldots'' and
``Geography of \ldots'' as separate categories), 
and users speaking a given language on a given level.

Moving on to the sizes of incoherent components,
we note that the largest in $\mathcal{A}$ (top middle panel in Figure \ref{fig:compsizes})
is well above the best-fit line ($\gamma \approx 0.587$).
This anomaly is an example of the so called ``king effect'',
discovered by Laherr\`ere \cite{ComptesRendus:1996:Laherrere}
while analyzing the sizes of the world's oilfields:
the largest element is much larger than
a log-log regression would predict.
The same component is a clear outlier in the distribution of the number
of English articles in components (top right panel in Figure \ref{fig:compsizes}).

To give a perspective: the largest component consists of 72\,284 articles, including 3\,184 articles in English,
while in theory each component should contain at most one article in English,
and its size should be bounded by the number of language editions, i.e.,\! about 250.
It contains articles on such a diverse subjects as:
``Abelian group'', ``Beekeeping'', ``Chinese poetry'', and ``Districts of Luxembourg''.
The second largest component in terms of the total number of articles contains 7\,004 nodes,
while the second largest in terms of the number of English articles contains 221 nodes.

Table \ref{tab:zipfsizes} presents the parameters of the best fits
corresponding to the lines in Figure \ref{fig:compsizes}.

\begin{table}
\caption{\label{tab:zipfsizes}The power law applied to the component sizes of the article and category networks.
``C'', ``I'' and ``E'' refer to the left, middle and right column in Figure \ref{fig:compsizes} (respectively).
It is tested whether the relation between a component size $s$ and its rank $r$ is indeed $s \sim r^{-\gamma}$.
The last column denotes adjusted $R^2$ -- a measure of correlation.}
\begin{ruledtabular}
\begin{tabular}{lcccc}
& \multicolumn{2}{c}{Range} & \multicolumn{2}{c}{Fit results} \\
& size & ranks & $\gamma$ & $R^2$ \\
\hline
Articles (C) & $[75, \infty)$  & $(-\infty, 2\,173]$ & 0.120 & 0.9852 \\
Articles (C) & $(-\infty, 75)$ & $(2\,173, \infty)$ & 0.544 & 0.9664 \\
Articles (I) & $[15, \infty)$ & $(-\infty, 23\,463]$ & 0.587 & 0.9889 \\
Articles (E) & $(-\infty, \infty)$ & $(-\infty, \infty)$ & 0.469 & 0.9690 \\
Categories (C) & $(-\infty, 27]$ & $[5\,638, \infty)$ & 0.196 & 0.9811 \\
Categories (C) & $(27, \infty)$ & $(-\infty, 5\,638)$ & 0.970 & 0.9855 \\
Categories (I) & $(-\infty, 20]$ & $[2\,472, \infty)$ & 0.493 & 0.9897 \\
\end{tabular}
\end{ruledtabular}
\end{table}

\begin{figure}
\begin{center}$
\begin{array}{cc}
\includegraphics[width=0.23\textwidth]{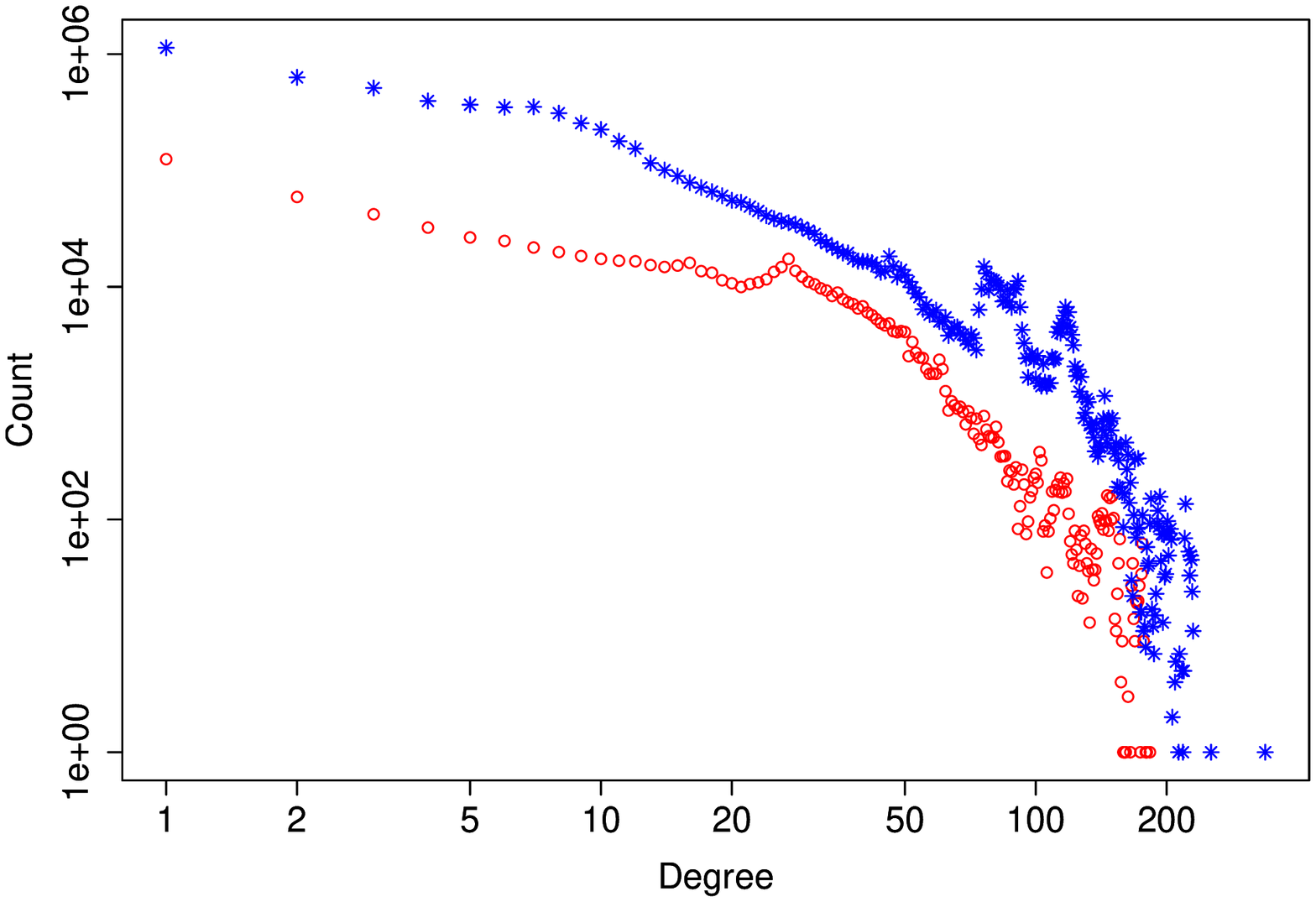} &
\includegraphics[width=0.23\textwidth]{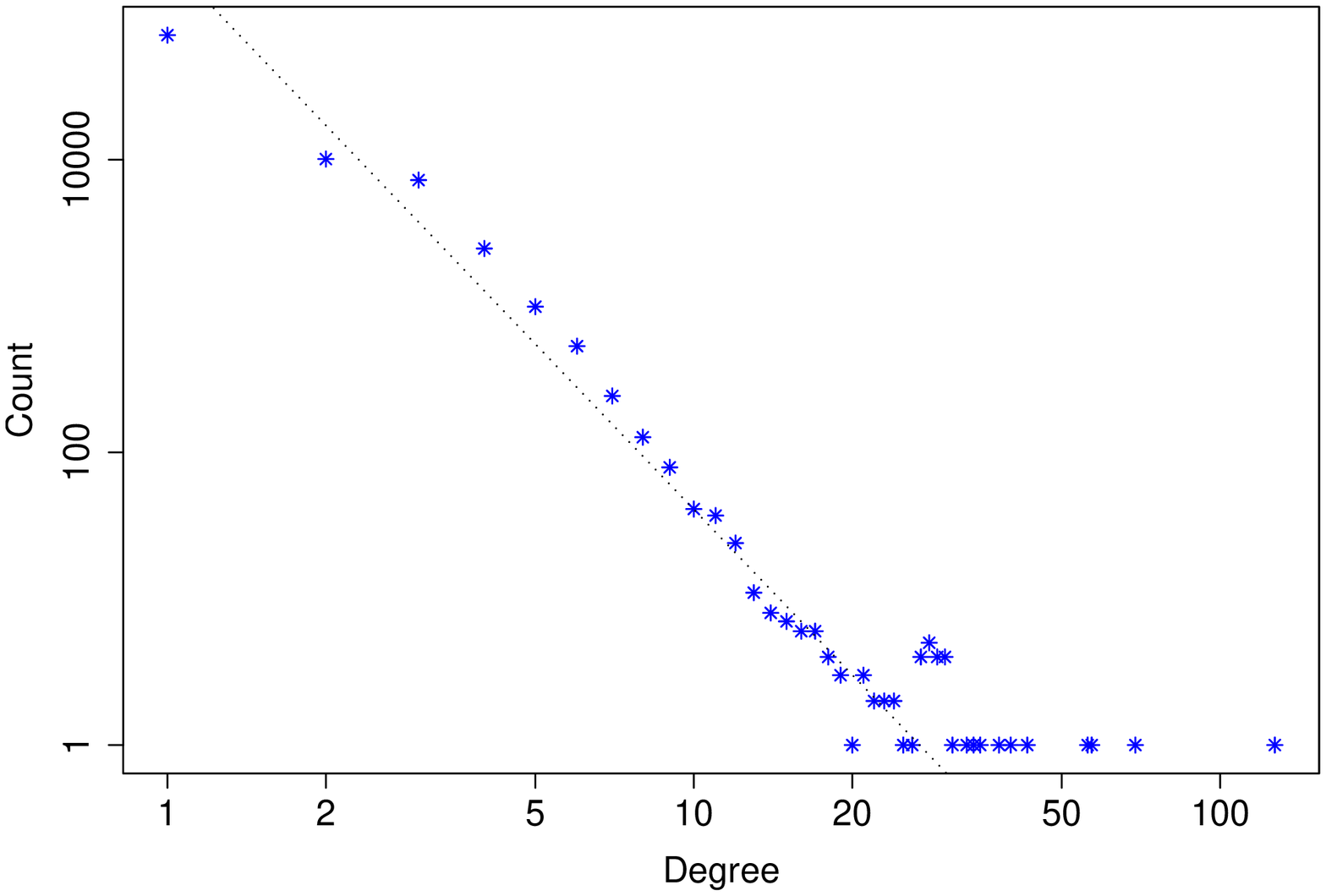} \\
\end{array}$
\end{center}
\caption{\label{fig:degree}%
(Left) Degree distribution for the article network (blue stars) and category network (red circles).
Log-log scale, degree 0 omitted.
The plateau at 74-93 and the peak at 117-119 in the case of articles are commented on in the text.
(Right) Degree distribution for the skeleton of the article network (log-log scale).
The skeleton extraction procedure is described in the text.
Only the incoherent components are accounted for, degree 0 is omited.
The distribution fits the power law with $\gamma \approx 3.75$ (adjusted $R^2 \approx 0.9770$).
The peak at 27-30 is commented on in the text.
}
\end{figure}

The left panel of Figure \ref{fig:degree} presents the node degree distribution in both researched networks
(note that both axes are logarithmic).
The median node degree is $6$ in $\mathcal{A}$ and $12$ in $\mathcal{C}$.
There are two anomalies in the distribution in the case of the article network:
a plateau spanning degrees 74-93, and a peak at degrees 117-119.
Both phenomena have plausible explanations.
The plateau is a result of articles where the subject is on years of the current era.
Some editions contain articles on all the 2000+ years,
others only on the more recent years.
For example, there are approx. 88 language editions covering year 1709,
82 covering year 1209 and 74 covering year 509.
The articles on a given year are usually forming a clique,
thus each article has degree equal to one less than the size of the clique.
As a consequence, we observe that an increased number of nodes with degrees 74-93
relates to a high number of cliques of sizes 75-94.
On the other hand, the peak at degrees 117-119, relates to
articles on days of the year.
There are approximately 120 language editions where such articles are present,
and these editions usually contain articles on all the 366 days.
Most of the groups of articles are connected in cliques,
hence a peak in the degree distribution.
The article with the highest degree (337) is
``ca:Llista de personatges de la Mitologia Eg\'{i}pcia''
which contains short descriptions of various gods of the Egyptian mythology.
A number of pages on articles in other languages, including 42 from the English edition,
contain interlanguage links to (redirects to) this page.

\begin{figure}
\begin{center}$
\begin{array}{cc}
\includegraphics[width=0.23\textwidth]{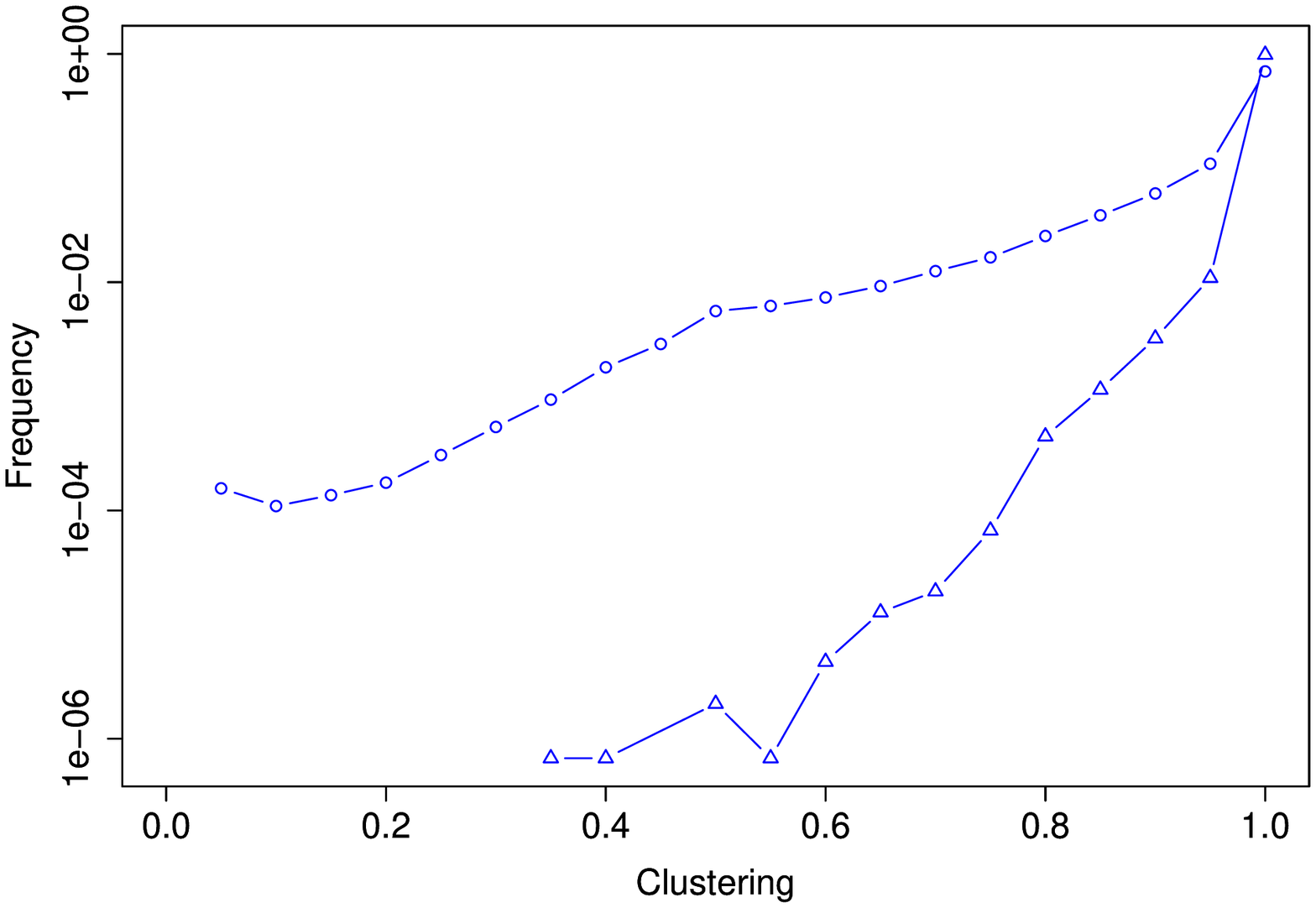} &
\includegraphics[width=0.23\textwidth]{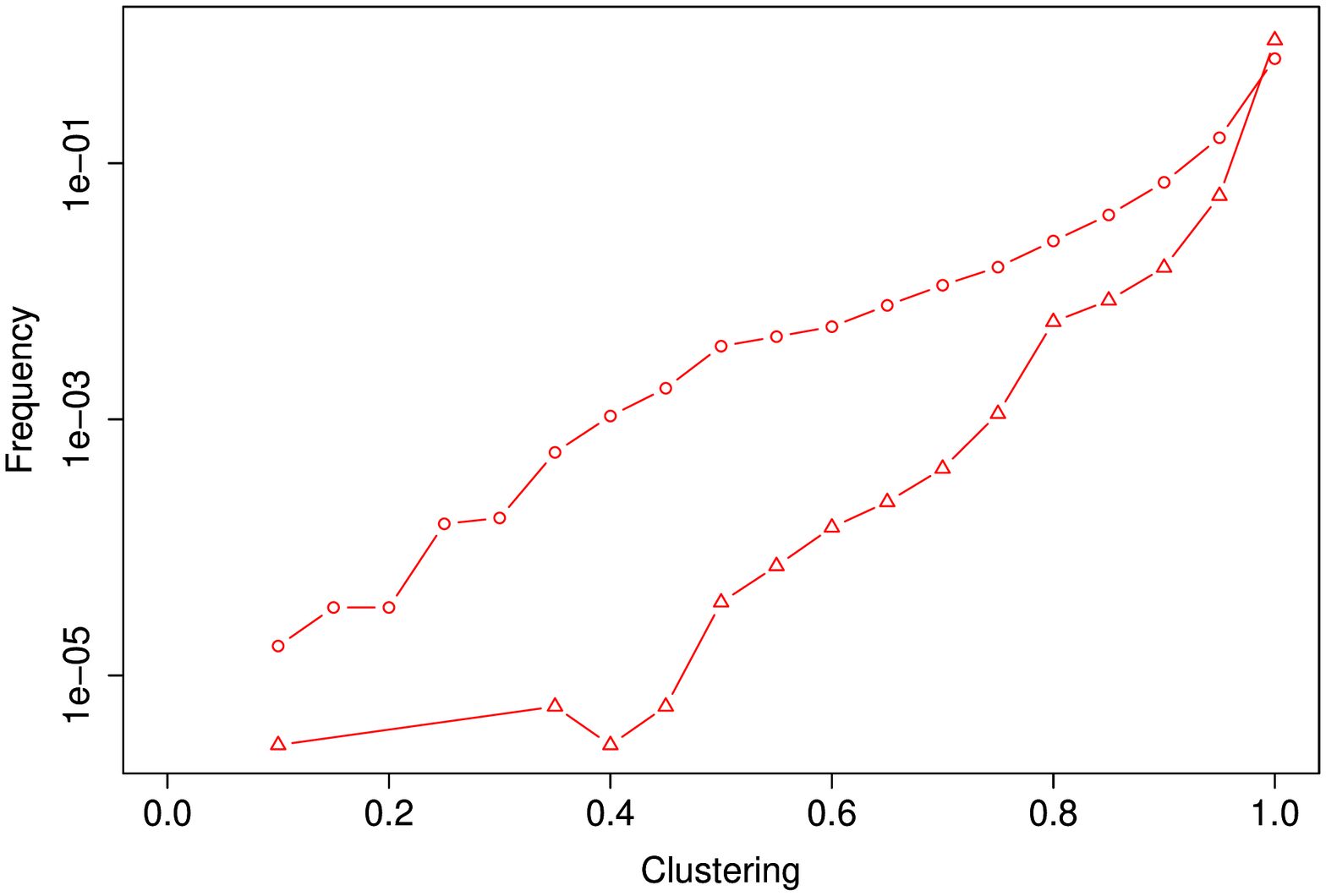} \\
\end{array}$
\end{center}
\caption{\label{fig:clustering} Distribution of the clustering coefficient values
of nodes.
Only the nodes with degree $> 9$ have been accounted for.
The left diagram presents distributions for the article network, the right diagram presents distributions for the category network.
Circles denote values for the incoherent components, triangles -- for the coherent ones.
Note that the y-axis is logarithmic, so the vast majority of nodes have clustering coefficient close to one.}
\end{figure}

Watts and Strogatz \cite{Nature:1998:Watts}
have demonstrated the usefulness of an indicator named
{\em clustering coefficient} in describing network topologies.
The clustering coefficient of a perfectly coherent and complete network
of interlanguage links would be $100\%$.
In reality, for both networks the value is quite high (approx. 97\%),
with over 98\% for the coherent components,
and approx. 91\% for the incoherent ones.
Figure \ref{fig:clustering} presents the distribution of the values
of the clustering coefficient for nodes having at least $10$ neighbors
\footnote{If the lower degrees were included, peaks at
$\frac{1}{2}$, $\frac{1}{3}$, $\frac{2}{3}$, $\dots$ would be visible,
since the low-degree nodes have only a few possible values of the clustering coefficient.
We have decided to ``subtract'' the expected peaks and show the less-obvious
pattern.}.
As expected, a low clustering coefficient is fairly uncommon.
More interesting is
the conditional probability that a node with degree $> 9$
and clustering coefficient $< 80\%$ will be part of an incoherent component:
it is approximately equal to 99.58\% in the case of $\mathcal{A}$ and ``only'' 77.48\% in the case of $\mathcal{C}$.

Let us summarize the results so far:
having analyzed the distributions of component sizes, degrees, and clustering coefficients,
we have found that the network of interlanguage links mainly consists of ``near-cliques''.
There are rare connections between the cliques, which are usually symptoms of incoherence.
Our next question is: what is the topology of these rare connections?

We would like to extract the ``skeleton'' of an incoherent component,
a network in which each set of nodes representing a given topic (usually a ``near-clique'')
is shrunk to a single point,
thus revealing the connections between separate topics.
Of course, partitioning a network into topics requires expert knowledge, which we cannot provide.
Instead, we propose a very simple method of extracting
an approximate structure of the skeleton network:
\begin{enumerate}
\item choose any of the most frequently occurring languages, the nodes in this language will be reference nodes;
\item for each node $v$: find the closest reference node(s) $z(v)$;
\item\label{it:skeleton-a} while there exists a pair of connected nodes $v_1$, $v_2$ such that
$z(v_1) = z(v_2)$ and $|z(v_1)| = |z(v_2)| = 1$: merge $v_1$ and $v_2$;
\item while there exists a pair of connected nodes $v_1$, $v_2$ s.t.
$|z(v_1)| > 1$ and $|z(v_2)| > 1$: merge $v_1$ and $v_2$;
\item while there exists a node $v$ such that
$|z(v)| = 2$ and $v$ is connected to exactly two other nodes:
remove $v$ and directly connect the two other nodes.
\end{enumerate}
Merging two nodes $v_1$ and $v_2$ means replacing them with a new node $v$,
connecting the new node with all the neighbors of $v_1$ and $v_2$,
and setting $z(v) := z(v_1) \cup z(v_2)$.

\begin{figure}
\begin{center}
\includegraphics[width=0.46\textwidth]{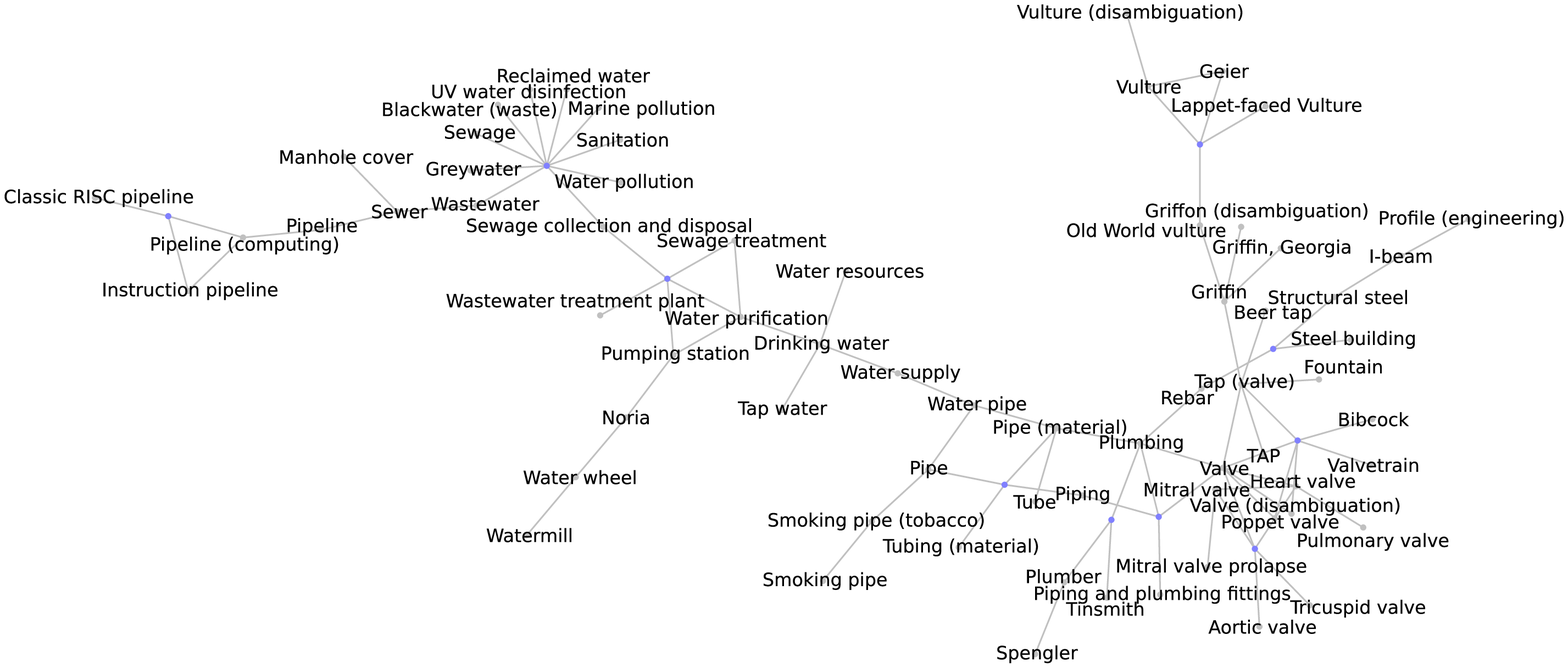}
\end{center}
\caption{\label{fig:pipes} Skeleton of a medium-sized incoherent component (812 articles, including 47 in English).
The skeleton extraction procedure is described in the text.}
\end{figure}

Figure \ref{fig:pipes} presents a result of the skeleton extraction procedure
applied to a middle-sized component.
After extracting the skeleton of the entire $\mathcal{A}$ network,
the average degree of a skeleton node is approx.\! 1.17,
and the clustering coefficient is approx. 37\%.
The distribution of node degrees is shown in the right panel of Figure \ref{fig:degree}.
The distribution is power-law (with $\gamma \approx 3.75$), indicating a scale-free network.
The peak at degrees 27-30 is yet another result of mass-edition,
this time related to articles on the days of the year (such as: {\em en:December 1}, {\em en:December 2}, etc.)
In 10 out of 12 cases, at least one language edition contains a bizarre copy-and-paste
error that connects all the days of a given month, for example
all the 30 articles on the days of September from the Hindu edition contain
an interlanguage link to the article in Kannada on September 11.
Thus, in the skeleton network, the node representing September 11 has 29 neighbors.
Note that in the ideal case (no incoherence) the skeleton network should consist solely of isolated nodes,
i.e., should contain no links at all.

Summing up, we have presented the surprisingly complex topology of the interlanguage links in Wikipedia.
Instead of a set of isolated cliques, the structure can be informally described as
a scale-free network of loosely interconnected near-cliques.
From a user's point of view, lack of coherence results in semantic drift,
e.g.\! {\it en:Pipeline} and {\it en:Vulture} are connected
by a series of interlanguage links which are supposed to model equivalence (cf. Figure \ref{fig:pipes}).

The results of our research motivated us to create a web service ({\tt http://wikitools.icm.edu.pl})
where Wikipedians may find detailed analysis of each incoherent component,
together with relevant edit recommendations.
We advertized the web service on the Wikipedia's mailing list,
which initiated a short discussion \cite{WikiEn:2008:ILL}.
Following the discussion, a Wikipedian changed
the wording of the definition of an interlanguage link
from ``to the same subject''
to ``to one or more nearly equivalent or exactly equivalent pages''.
However, the topology described in this paper indicates serious incoherence
even under the slightly relaxed definition.

\bibliography{ill-topology}

\end{document}